\documentclass[10pt]{article}
\usepackage{amsfonts}
\usepackage{amssymb,amsmath}
\def\AP{{\it Ann. Phys.}}
\def\NP{{\it Nucl. Phys.}}

\def\PL{{\it Phys. Lett.}}
\def\PR{{\it Phys.  Rev.}}
\def\CQG{{\it Class.  Quan.  Grav.}}

\def\MPL{{\it Mod. Phys. Lett.}}
\def\JHEP{{\it J.High\ Energy\ Phys.}}

\newcommand{\mat}[1]{\left[\begin{matrix}#1\end{matrix}\right]}
\newcommand{\cor}[1]{\left\{#1\right\}}

\begin{document}
\begin{titlepage}

\hfill{Preprint {\bf SB/F/05-330}} \hrule \vskip 2.5cm
\centerline{\bf On the Quantization of Massive Superparticles}
\vskip 1cm

\centerline{N. Hatcher}
\begin{description}
\item[]{\it  Universidad Sim\'on Bol\'{\i}var, Departamento de Matem{\'a}ticas, Apartado Postal 89000,
Caracas 1080-A, Venezuela.}
\end{description}
\centerline{A. Restuccia and J. Stephany} \vskip 4mm
\begin{description}
\item[]{\it  Universidad Sim\'on Bol\'{\i}var, Departamento de F\'{\i}sica, Apartado Postal 89000,
Caracas 1080-A, Venezuela.}
\item[]{\it \ \ \ e-mail:  nhatcher@fis.usb.ve, arestu@usb.ve,
 stephany@usb.ve}
\end{description}
\vskip 1cm

{\bf Abstract}

\vskip 3mm

\noindent We consider the action of the $D=11$ supermembrane
wrapping a compactified sector of the target space  in such a way
that a non trivial central charge in the SUSY algebra is induced.
 We show that the dynamics of the center of mass
corresponds to a superparticle in $D=9$ with additional fermionic
terms associated to the  central charges . We perform the
covariant quantization of this system following a direct approach
which introduces an equivalent action for the system which has
only first class constraints allowing to obtain the space of
physical states in a covariant way. The resulting multiplet
contains $2^8$ states corresponding to a $KKB$ ultrashort
multiplet.

\vskip 2cm
\hrule
\bigskip
\centerline{\bf UNIVERSIDAD SIMON BOLIVAR}
\vfill
\end{titlepage}

\section{Introduction}
The theory of the supermembrane \cite{Bergshoeff} is a key element
in the intricate network of related systems which are expected to
define the consistent non perturbative aspects of
 quantum superstrings.
It was originally constructed as the $D=11$ extended object which
propagates consistently on a supergravity background but its role
in $M$-theory turn out to have more implications. It was shown in
\cite{deWitLN} that when the supermembrane is immersed in a $D=11$
Minkowski space its spectrum is continuous from $[0,\infty]$ but
it is still unknown if there is a massless sector corresponding to
the $D=11$ supergravity. It was also argued in Ref. \cite{deWitPP}
that the supermembrane on a compact target space should also have
a continuous spectrum. In contrast the spectrum of the $D=11$
supermembrane wrapping a two cycle when the configuration space
satisfies the topological condition which implies a non trivial
central charge in the supersymmetric algebra, was shown to be a
discrete set with finite multiplicity \cite{BoultonGR}.

In this paper we  focus in the low energy spectrum of the
supermembrane wrapped on a two cycle by considering the covariant
quantization of the associated superparticle. As we show below,
the resulting action describes a massive superparticle with an
additional spinorial term arising from the non trivial wrapping of
the supermembrane on the two-cycle. In the massless case, as is
well known, the covariant quantization of the
Casalbuoni-Brink-Schwarz superparticle in $D=10$ \cite{Brink}
present similar obstacles related to the mixing of first and
second class constraints associated with the
$\kappa$-symmetry\cite{Siegel}, as the Green-Schwarz superstring.
The covariant gauge fixing of the symmetry was performed in
\cite{Kallosh,Diaz,Restuccia} by introducing an infinite tower of
auxiliary fields . More recently, a new formulation  was advanced
in terms of pure spinors which requires a finite number of
fields\cite{Berkovits}. The dynamics of a superparticle with a
central charge was first considered in \cite{AzcarragaL,Aldaya}
and later in \cite{Huq}. However the central charges arose in
these cases from considerations different to those presented here.
The covariant quantization of this superparticle differs from the
usual massive superparticle since the central charge implies the
existence of a $\kappa$-symmetry additionally to the second class
constraints already present for the massive superparticle. Several
formulation to quantize the superparticle with central charges
were proposed \cite{Huq,FrydryszakL,Yugo}. For example in
\cite{Yugo} a $BRST$ charge was constructed also from a infinitely
reducible set of first class constraints. Fortunately as we
discuss in section 4 the structure of the system allows  a direct
approach in which the physical degrees of freedom are identified
in a covariant way.

\section{The Supermembrane in $M^9\times \Sigma$}

The supermembrane action immersed in $D=11$ target space was
obtained originally in Ref. \cite{Bergshoeff} and is given by,
\begin{eqnarray}
I=-\frac{1}{8\pi^2}\int d^3\xi \left[\sqrt{-g}g^{ij} \pi_i^{\mu}
\pi_{j\mu}-\sqrt{-g}
+i\epsilon^{ijk}\pi_i^{\mu} \pi_{j}^{\nu}\bar{\psi}\Gamma_{\mu\nu}\partial_k\psi\right]\nonumber \\
+\left[
\epsilon^{ijk}\pi_i^{\mu}\bar{\psi}\Gamma_{\mu\nu}\partial_j\psi\bar{\psi}\Gamma_{\nu}\partial_k\psi-
\frac{i}{3}\epsilon^{ijk}\bar{\psi}\Gamma_{\mu\nu}\partial_i\psi\bar{\psi}\Gamma_{\nu}\partial_j\psi
\psi\bar{\psi}\Gamma_{\nu}\partial_k\psi\right]\ \ ,
\end{eqnarray}
where $\psi$ is a Majorana spinor and
\begin{equation}
\pi^{\mu}_i=\partial_iX^{\mu}-i\bar\psi\Gamma^{\mu}\partial_i\psi
\end{equation}
with $X^{\mu}$ the space time coordinate of the membrane. The
supermembrane tension has been fixed to 1.

We are interested in the case when the $D=11$ target space has a
compactified sector admitting a minimal immersion from the base
manifold $\Sigma\times R$ into it. $\Sigma$  is a Riemann surface
of genus $g>0$ and $R$ corresponds to the range of the time
variable $\tau$.

To be specific let us consider the case in which $\Sigma$ is a
torus and the compactified target space is $M^9\times S^1\times
S^1$ \cite{BoultonGMR,BoultonGR,GarciaR}, although more general
compactified target spaces for which there exists a minimal
immersion from $\Sigma$ into the target space may also be treated
along the lines we follow here \cite{BellorinR}.

The minimal immersion is constructed from the harmonic one forms
over $\sigma$ denoted by $d\hat x^r$, $r=1,2$. We consider a pair
of harmonic one-forms over $\Sigma$ satisfying
\begin{equation}
\oint_{C_s}d\hat x^r=2\pi m^r_s \ \ \ ,\ \ \ \ \ \ \ \ r,s=1,2
\end{equation}
where $m^r_s$ are integers and $C_j$ is a basis of the homology on
$\Sigma$, together with
\begin{equation}
\label{area} \int_{\Sigma}d\hat x^r\wedge d\hat x^s=n
A(\Sigma)\neq 0 \ \ \ ,
\end{equation}
where $n=det\,m^r_s$ and $A(\Sigma)$ is the area of $\Sigma$. The
first condition implies that each $\hat x^r$ may define a map over
a circle $S^1$. It also implies the equality in (\ref{area}). The
condition $A(\Sigma)\neq 0$ is the non trivial part of this
relation.

The map from $\Sigma$ onto $S^1 \times S^1$ is given, with $P_0$ a
fixed point in $\Sigma$,  by
\begin{equation}
P\in \Sigma \rightarrow \int_{P_0}^{P}d\hat x^r \ \ \  {\rm mod}\
(2\pi n^r) \  ,\ \ r=1,2\ \ .
\end{equation}

Condition (\ref{area}) implies that the algebra of the
supermembrane has a non-trivial central charge
$Z^{rs}=\epsilon^{rs}n A(\Sigma)$ with $n$ the number of times the
supermembrane wraps $S^1 \times S^1$. It may be shown that the
above map is a local minimum of the hamiltonian of the
supermembrane \cite{OvalleMR} and defines a minimal immersion from
$\Sigma$ to $S^1 \times S^1$ \cite{BellorinR}. It is a solution of
the supermembrane field equations which preserves one half of the
original supersymmetry. The most general configuration space and
hamiltonian for the supermembrane with the above base manifold and
target space with a non-trivial central charge was obtained in
\cite{BoultonGR,GarciaR,BoultonGMR} and shown to have a discrete
spectrum. We are interested here in the quantization of the
corresponding superparticle with non-trivial central charges.

\section{Compactification of the Supermembrane}
We consider now the covariant quantization of the center of mass
of the supermembrane which corresponds to a superparticle. To do
so we restrict the configuration space by the following
conditions,
\begin{eqnarray}
X^m&=&X^m(\tau) \ \ \ ,\ \ \ m=0,\ldots ,8\\
\psi &=& \psi(\tau)\ \ \ ,\\
X^{r+8} &=& \hat x^r(\sigma)\ \ \ ,\ \ \ r=1,2
\end{eqnarray}
where $\sigma^a , a=1,2$ denote local coordinates over $\Sigma$.
On this class of configurations the supermembrane action reduces
to
\begin{gather}
S \rightarrow k\int d\tau \left(e^{-1}\omega^m \omega_m -e - i
\psi^{\beta}(\frac{1}{2}C\Gamma_{rs}\epsilon^{rs})_{\beta
\gamma}\dot\psi^\gamma\right)\ \ \ ,\nonumber \\ \omega^m=\dot
X^m-i\bar\psi\Gamma^m\dot\psi\ \ .
\end{gather}
where $C$ is the charge conjugation matrix in $D=11$ and
$\Gamma_{rs}$ is the antisymmetric product
$\frac{1}{2}(\Gamma_r\Gamma_s-\Gamma_s\Gamma_r)$.(The conventions
for the $\Gamma$ matrices are given in the appendix.
$C\Gamma_{rs}$ is symmetric on the spinorial indices). The
constant $k$ is given by $k=\frac{nA(\Sigma)}{8\pi^2}$.

 In the following section we
perform the covariant quantization of this system following a
direct approach by  introducing a new action for the superparticle
with central charges with first class constraints only, that
allows to obtain the space of physical states in a covariant way.

We use now a particular representation of the Dirac matrices. We
consider $\gamma^m,\  m=0, \ldots, 8$ the set of Dirac matrices in
$D=9$ satisfying \mbox{$\{\gamma^{\mu},\gamma^{\nu}\}=-2\eta^{\mu
\nu}$}. We take as in \cite{Huq},
\begin{equation}
\Gamma^m=\left[{\begin{array}{cc}
0 & i\gamma^m \\
-i \gamma^m & 0
\end{array}}
\right]\ \ \ ,\ \ \ \Gamma^9=\left[{\begin{array}{cc}
0 & i \\
i  & 0
\end{array}}
\right]\ \ \ .
\end{equation}
Then we have \mbox{$\{\gamma^{\mu},\gamma^{\nu}\}=-2\eta^{\mu
\nu},\  \mu,\nu=0, \ldots, 9,11$}, with
\begin {equation}
\Gamma^{11}=\Gamma^0\Gamma^1\ldots\Gamma^9=\left[{\begin{array}{cc}
\mathbb{I} & 0 \\
0  & -\mathbb{I}
\end{array}}
\right]\ \ .
\end{equation}
The charge conjugation matrix in $D=11$ in terms of the
corresponding one in $D=9$ is (see the appendix)
\begin{equation}
C=\left[{\begin{array}{cc}
0 & -i\tilde C \\
i \tilde C & 0
\end{array}}
\right]\ \ \ ,\label{CtoC}
\end{equation}
where $C^T=-C$ and $\tilde C^T=\tilde C$. Now we decompose the
$D=11$ spinors in terms of the $D=9$ ones
\begin{equation}
\psi=\left({\begin{array}{c}
\theta_1 \\
\theta_2
\end{array}}
\right)\ \ \ .
\end{equation}
The Majorana condition $\bar\psi=-\psi^TC$ in $D=11$ implies
\begin{equation}
\bar\theta_A=\theta_A^T\tilde C \ \ ,
\end{equation}
where $\bar\theta_A=\theta_A^\dagger\gamma^0$. The action for the
superparticle with central charges reduces then to

\begin{equation}
S=k\int d\tau \left(e^{-1}\omega^m\omega_m -e
-i\bar\theta_A\dot\theta_A\right)
\end{equation}
where now
\begin{equation}
\omega^m=\dot X^m-i\bar\omega_A\gamma^m\dot\omega_A \ \ .
\end{equation}

\section{The Hamiltonian for the Superparticle with central charges}

Let us introduce here the mass parameter $m=2k$ and the
projectors,
\begin{equation}
P\pm=\frac{1}{2m}(m\pm\gamma^mp_m)\ \ ,
\end{equation}
which satisfy
\begin{eqnarray}
P_+P_+&=&-\frac{1}{4m^2}(p^2+m^2)+P_+\ ,\ \ \ P_-P_-=-\frac{1}{4m^2}(p^2+m^2)+P_-\ \ ,\nonumber \\
 P_+P_-&=&P_-P_+= \frac{1}{4m^2}(p^2+m^2)\ ,\ \ \ P_++P_-=\mathbb{I}\ \ .
\end{eqnarray}
The conjugate momenta to $X^m$ are
\begin{equation}
p_m=2ke^{-1}\omega_m\ ,
\end{equation}
and the conjugate momenta to  $\theta_A$, with $m=2k$ are
\begin{equation}
\bar\pi_A=2im\bar\theta_AP_-
\end{equation}
or equivalently
\begin{equation}
\pi_A=-2imP_-\theta_A\ \ .
\end{equation}
Introducing the Lagrange multiplier $\lambda\equiv\frac{e}{4k}$,
the hamiltonian may then be expressed as
\begin{equation}
\mathcal{H}=\lambda(p^2+m^2)
\end{equation}
subject to
\begin{equation}
\label{mixture} \pi_A+2imP_-\theta_A=0 \ \ ,
\end{equation}
which are a mixture of first and second class constraints.

The set of constraints (\ref{mixture}) together with the mass
shell condition $p^2+m^2=0$ are equivalent to
\begin{eqnarray}
p^2+m^2=0 \label{IIIa}\ \ ,\\
P_+\pi_A=0 \label{IIIb}\ \ ,\\
P_-\pi_A+2imP_-\theta_A=0\ \ .\label{IIIc}
\end{eqnarray}
Now (\ref{IIIa}) and (\ref{IIIb}) are first class constraints
while (\ref{IIIc}) is a mixture of first and second class
constraints.

We may then consider the set,
\begin{eqnarray}
p^2+m^2=0 \label{IVa}\\
\varphi_1=P_+(\pi_1+i\pi_2)=0  \label{IVb}\\
\ \ \varphi_2=P_+(\pi_1-i\pi_2)=0 \label{IVbb}\\
\Omega_1=P_-(\pi_1+i\pi_2)+2imP_-(\theta_1+i\theta_2)=0\label{IVc} \\
\Omega_2=P_-(\pi_1-i\pi_2)+2imP_-(\theta_1-i\theta_2)=0
\label{IVd}
\end{eqnarray}
with the non trivial bracket,
\begin{eqnarray}
\{\Omega_1,\Omega_2\}=-4imP_-\ \ .
\end{eqnarray}

We also note that the symplectic terms in the canonical action
$\bar\pi_A\dot\theta_A$ may be decomposed as
\begin{eqnarray}
\bar\pi_A\dot\theta_A=
\frac{1}{2}{\overline{(\pi_1+i\pi_2)}}(\dot\theta_1-i\dot\theta_2)+
\frac{1}{2}{\overline{(\pi_1-i\pi_2)}}(\dot\theta_1+i\dot\theta_2)
\ \ ,
\end{eqnarray}
and we identify the conjugate pairs
$(\pi_1+i\pi_2)$,$(\dot\theta_1-i\dot\theta_2)$ and
$(\pi_1-i\pi_2)$,$(\dot\theta_1+i\dot\theta_2)$. We notice here
that the pair $\Omega_1$ and $\Omega_2$ may be regarded as a first
class constraint ($\Omega_1$) and  an associate gauge fixing
condition ($\Omega_2$). The contribution of this pair to the
functional measure when taken as a pair of second class
constraints or in the way proposed is exactly the same
\cite{Gianvittorio}, \cite{Restuccia2}. With this observation we
can finally define our system as a gauge system restricted by the
set of first class constraints (\ref{IVa}), (\ref{IVb}) and
(\ref{IVc}). If one then would like to impose (\ref{IVd}) as a
partial gauge fixing condition the original set of constraints is
recovered, but there is freedom to impose a different set of
admissible gauge fixing conditions since the functional integral
is invariant to this choice (In this case there are no gauge
anomalies).

To continue we consider the partial gauge fixing conditions,
\begin{eqnarray}
P_+(\theta_1-i\theta_2)=0 \ \ ,\\
P_-(\theta_1-i\theta_2)=0 \ \ ,
\end{eqnarray}
corresponding to the symmetry generated by $\varphi_1$ and
$\Omega_1$ respectively. This gauge fixing condition contributes
to the functional integral with a constant factor independent of
the fields. The canonical variables $(\pi_1 +i\pi_2)$ and
$(\theta_1-i\theta_2)$ may then be integrated out  from the
functional integral.

We are thus left with the canonical action,

\begin{eqnarray}
\mathcal{L}&=&p\dot X+\bar\pi_A\dot\theta_A-\mathcal{H}\\
&=&p\dot X +
\frac{1}{2}\overline{(\pi_1-i\pi_2)}(\dot\theta_1+i\dot\theta_2)\
\ ,
\end{eqnarray}
constrained by (\ref{IVa}) and  (\ref{IVbb}). Let us introduce the
variables $\tilde\theta=\theta_1+i\theta_2$ and
$\tilde\pi=\frac{1}{2}(\pi_1-i\pi_2)$. Notice that $\tilde\theta$
is a complex spinor which does not satisfy the pseudo Majorana
condition $\bar\theta=\theta^T\tilde C$

We  may finally consider the partial gauge fixing condition
\begin{equation}
P_+(\theta_1+i\theta_2)=0,
\end{equation}
and perform a canonical reduction to
\begin{eqnarray}
\mathcal{L}=p\dot X + \overline{(P_-\tilde \pi)}(P_-\tilde
\theta)\ \ ,
\end{eqnarray}
subject to the mass shell condition (\ref{IIIa}). The space of
physical states is obtained by considering superfields depending
in $P_-\tilde\theta$ and not on its complex conjugate. Since we
have that $P_-+P_+=1$ without using $p^2+m^2=0$ the degrees of
freedom in $P_-\tilde\theta$ are exactly half of the ones in
$\tilde\theta$. They are therefore $2^8$ bosonic and fermionic
on-shell degrees of freedom. It corresponds as we discuss below to
a $D=9$ $KKB$ multiplet.

\section{Conclusion}
We have covariantly quantized the $D=9$ superparticle associated
to the $D=11$ supermembrane wrapped on a torus with a non trivial
central charge.\\
The Hilbert space of states we obtained is described in terms of
an on-shell superfield $\Psi(x^\mu,\tilde{\theta}_-)$. The
spinorial variable $\tilde{\theta}_-$ has 8 independent variables.
The superfield has $2^8$ degrees of freedom that fit neatly in a
$D=9$ massive supermultiplet with central charge. The general form
of this central charge in $D=9$ arising from the supermembrane
algebra in $D=11$ is \cite{Nicolai}
\begin{eqnarray}
Z^{ij}=Z\delta^{ij}-(P_9\sigma^3-P_{10}\sigma^1) \ \ ,
\end{eqnarray}
with a BPS mass given by
\begin{eqnarray}
M=\sqrt{P_9^2+P_{10}^2}+|Z|\ \ .
\end{eqnarray}
In this paper we have taken $P_9=P_{10}=0$. The multiplet that we
have obtained from the quantization of the \mbox{$D=9$}
superparticle  corresponds to a
ultrashort KKB supermultiplet \cite{Nicolai}.\\
Due to the nature of our central charge we are studying the
winding modes of the supermembrane on a torus and neglecting the
Kaluza-Klein modes. As is well known \cite{Schwarz} this states
should correspond to the Kaluza-Klein modes of the IIB superstring
wrapped on $S^1$. Our results confirm this correspondence.

\section{Acknowledgments} This work was supported
by Did-Usb grants Gid-30 and Gid-11 and by Fonacit grant
G-2001000712.

\appendix
\section{Appendix:Dirac matrices and Central charges}
We collect here some useful results about the supersymmetry
algebra in $D=9$. We take the "mostly plus" signature with the
Dirac matrices satisfying
$\{\Gamma^\mu,\Gamma^\nu\}=-2\eta^{\mu,\nu}$. We construct these
matrices recursively. Let $\gamma^\mu$ be a set of Dirac matrices
in $D-1$ dimensions ($D$ even), then in $D$ dimensions we have,
\begin{equation} \Gamma^\mu=\left[{\begin{array}{cc}
0 & S^\mu \\
\bar{S}^\mu & 0
\end{array}}
\right]\ \ \ ,\ \ \ {\begin{array}{cc}
S^0=\bar{S^0}=\mathbb{I} & S^i=\gamma^i\gamma^0 \\
\bar{S^i}=-S^i=\mathbb{I} & S^{D-1}=\gamma^0
\end{array}}\ \ .
\end{equation}
In an even dimensional space there exist matrices $B, \tilde{B},
C$ and $\tilde {C}$ such that
\begin{equation} {\begin{array}{ccc}
B\Gamma^\mu B^{-1}=-(\Gamma^{\mu})^*&\tilde{B}\Gamma^\mu \tilde{B}^{-1}=(\Gamma^{\mu})^* & B=\tilde{B}W \\
C\Gamma^\mu C^{-1}=-(\Gamma^{\mu})^T&\tilde{C}\Gamma^\mu
\tilde{C}^{-1}=(\Gamma^{\mu})^T &C=\tilde{B}\Gamma^0
\end{array}}
\end{equation}
with $W=i^{[D/2+1]}\Gamma^0....\Gamma^D$. In odd dimensions there
exist either $(B,C)$ or $(\tilde{B},\tilde{C})$. There exist also
a matrix $\tilde{B}_9$ in $D=9$ and a matrix ${B}_11$ in $D=11$
defined by,
\begin{equation}
\tilde{B}_9=\gamma^0\gamma^1\gamma^3\gamma^5\gamma^7 \ \ , \ \ \ \
{B}_{11}=\Gamma^2\Gamma^4\Gamma^6\Gamma^8\ \ ,
\end{equation}
such that
\begin{equation}
B_{11}=\left[{\begin{array}{cc}
\tilde{B}_9 & 0 \\
0 & -\tilde{B}_9
\end{array}}
\right]\ \ \ ,\ \ \ \ C_{11}=\left[{\begin{array}{cc}
0 & \tilde{C}_9\gamma^0 \\
-\tilde{C}_9\gamma^0  & 0
\end{array}}
\right]\ \ \ .
\end{equation}
To make contact with our notation of section 3, call that set of
matrices $\Gamma^\mu_H$ they are related to $\Gamma^\mu$ by a
unitary transformation
\begin{equation}
U\Gamma^\mu_H U^\dag=\Gamma^\mu\qquad U=\mat{i & 0\\ 0 &\gamma^0}\
\ .
\end{equation}
From here follows directly equation (\ref{CtoC}).\\
The most general supersymmetry algebra in $D=9$ and $N=2$ with
Lorentz invariant central charges is
\begin{equation}
\cor{Q_{ai},Q_{bj}}=2\delta{ij}p_\mu(\gamma^\mu
\tilde{C}^{-1})_{ab}+Z_{ij}\tilde{C}_{ab}\ \ ,
\end{equation}
with $Z_{ij}$ a real symmetric matrix. Note that we can write
\begin{eqnarray}
\gamma^0\tilde{C}=\mat{0 & J\\ J & 0},\qquad \tilde{C}=\mat{J &
0\\ 0 & J},\qquad J^T=J,\, J^2=I\ \ .
\end{eqnarray}
Since these two matrices commute they can be simultaneously
diagonalized. Then the algebra in the rest frame takes the form
\begin{eqnarray}
\cor{Q_{ai},Q_{bj}}=2m\delta_{ij}\mat{J & 0\\ 0 & -J}+Z_{ij}\mat{J
& 0\\ 0 & J}\ \ .
\end{eqnarray}
The algebra of the 32 supercharges splits into 4, $8\times 8$
blocks. In our case $Z_{ij}=2m\delta_{ij}$ and we find
\begin{eqnarray}
\cor{Q_{ai},Q_{bj}}=4m\delta_{ij}\mat{J & 0\\ 0 & 0}\ \ .
\end{eqnarray}
The entire representation may be obtained now as usual. We notice
that half of the supersymmetries are not present and the other
half build a representation of $2^8$ states.

\end{document}